\documentclass[12pt,a4paper,twoside]{article}
\usepackage{epsfig}
\usepackage{baltlat1}
\usepackage{wrapfig}
\begin{document}
\ \
\vspace{0.5mm}

\setcounter{page}{1}
\vspace{8mm}

\titlehead{Baltic Astronomy, vol.12, XXX--XXX, 2003.}

\titleb{THE REDSHIFT OF LONG GRBS'}

\begin{authorl}

\authorb{Z. Bagoly}{1} and
\authorb{I. Csabai}{2} and
\authorb{A. M\'esz\'aros}{3} and
\authorb{P. M\'esz\'aros}{4} and
\authorb{I. Horv\'ath}{5} and
\authorb{L. G. Bal\'azs}{6} and
\authorb{R. Vavrek}{7}
\end{authorl}

\begin{addressl}
\addressb{1}{Lab. for Information Technology, E\"{o}tv\"{o}s University, H-1117 Budapest, P\'azm\'any P. s.  1./A, Hungary}

\addressb{2}{Dept. of Physics for Complex Systems, E\"{o}tv\"{o}s University, H-1117 Budapest, P\'azm\'any P. s.  1./A, Hungary}

\addressb{3}{Astronomical Institute of the Charles University, V Hole\v{s}ovi\v{c}k\'ach 2, CZ-180 00 Prague 8, Czech Republic}

\addressb{4}{Dept. of  Astronomy \& Astrophysics, Pennsylvania State University, 525 Davey Lab., University Park, PA 16802, USA}

\addressb{5}{Dept. of Physics, Bolyai Military University, H-1456 Budapest, POB 12, Hungary}

\addressb{6}{Konkoly Observatory, H-1505 Budapest, POB 67, Hungary}

\addressb{7}{Max-Planck-Institut f\"{u}r Astronomie, D-69117 Heidelberg, 17 K\"{o}nigstuhl, Germany}
\end{addressl}

\submitb{Received October 20, 2003}

\begin{abstract}
The low energy spectra of some gamma-ray bursts' show excess components beside
the power-law dependence. The consequences of such a feature allows to estimate
the gamma photometric redshift of the long gamma-ray bursts in the BATSE
Catalog.  There is good correlation between the measured optical and the
estimated gamma photometric redshifts.  The estimated redshift values for the
long bright gamma-ray bursts are up to {\footnotesize $z=4$}, while for the the
faint long bursts - which should be up to {\footnotesize $z=20$} - the
redshifts cannot be determined unambiguously with this method. The redshift
distribution of all the gamma-ray bursts with known optical redshift agrees
quite well with the BATSE based gamma photometric redshift distribution.
\end{abstract}


\resthead{The Redshift of Long GRBs'}{Z.~Bagoly {\em et. al}}

\sectionb{1}{INTRODUCTION}

In this article we present a new method called {\it gamma photometric redshift
} (GPZ) estimation of the estimation of the redshifts for the long GRBs.  We
utilize the fact that broadband fluxes change systematically, as characteristic
spectral features redshift into, or out of the observational bands.  The
situation is in some sense similar to the optical observations of galaxies,
where for galaxies and quasars the photometric redshift estimation (Csabai {\em et. al} (2000), Budav\'ari {\em et. al} (2001))
achieved a great success in estimating redshifts from photometry only.  

We construct our {\em template spectrum} that will be used in the GPZ process
in the following manner: let the spectrum be a sum of the Band's function  and
of a low energy soft excess power-law function, observed in several cases
(Preece {\em et.  al} (2000)).  The low energy
cross-over is at $E_{cr}= 90$ keV,  $E_o=500$ keV, and the spectral indices are
$\alpha=3.2$, $\beta=0.5$ and $\gamma=3.0$.

Let us introduce the {\em peak flux ratio} (PFR hereafter) in the following way:
\begin{equation}
\mbox{PFR} = {{l_{34}-l_{12}} \over {l_{34}+l_{12}}}
\end{equation}
where $l_{ij}$ is the BATSE DISCSC flux in energy channel $E_i < E < E_j$,
here $E_1=25$ keV, $E_2=E_3=55$ keV, $E_4=100$ keV.

\vskip1mm
\begin{wrapfigure}{i}[0pt]{61mm}
\centerline{\psfig{figure=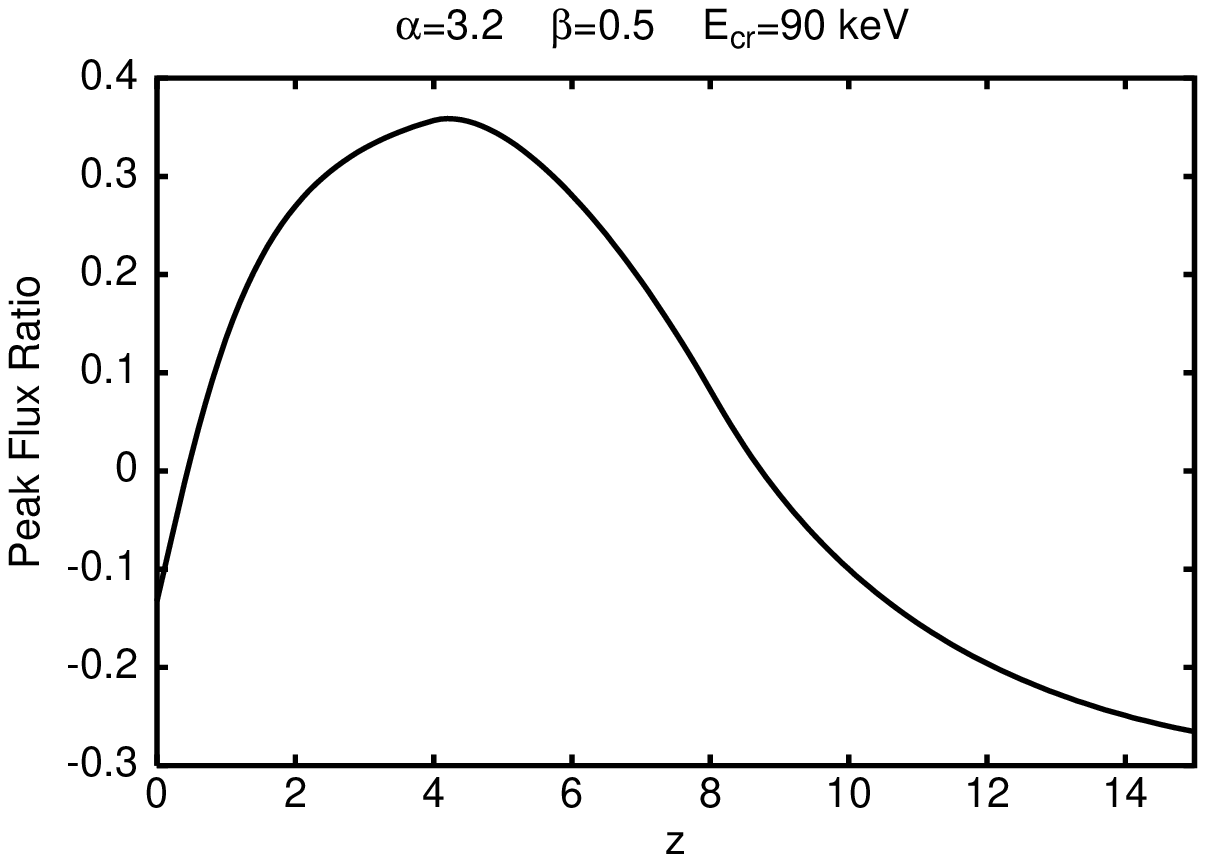,width=60truemm,angle=0,clip=}}
\captionb{1}{The theoretical PFR curves
calculated from the template spectrum using the
average detector response matrix.}
\end{wrapfigure}

The spectra are changing quite rapidly with
time; the typical timescale for the time variation is $\simeq (0.5 - 2.5)$~s
(Ryde \& Svensson (1999, 2000)).
Therefore, we will consider the spectra in the $320$~ms  time interval
centered around the peak-flux.  
If we redshift the template spectrum and use the 
detector response matrix of the given burst, 
we can get for any redshift the observed flux and the PFR value.

On Fig.~1. we plot the theoretical PFR curves calculated from the above defined
template spectrum using the {\em average} detector response matrices for the 8
bursts that have both BATSE data and measured redshifts (Klose (2000))
In the used range of $z$ (i.e. for $z ^{<}_{\sim} 4$)
the relation between $z$ and PFR is invertible, hence we can use it to estimate
the {\em gamma photometric redshift} (GPZ) from a measured PFR.  For the 7
considered GRBs (leaving out GRB associated with the supernova and GRB having
upper redshift limit only) the estimation error between the real $z$ and the
GPZ is $\Delta z= \approx 0.33$.

\sectionb{2}{ESTIMATION OF THE REDSHIFTS}

Here restrict ourselves to long and not very faint GRBs with $T_{90} > 10\;s$
and $F_{256} < 0.65$ photon/(cm${}^2$s) to avoid the problems with the
instrumental threshold (Pendleton {\em et. al}  (1997), Hakkila {\em et. al},
(2000)).  Introducing an another cut at $F_{256} > 2.00$ photon/(cm${}^2$s) we
can investigate roughly the brighter half of this sample.

As the soft-excess range redshifts out from the BATSE DISCSC energy channels
around $z \approx 4$, the theoretical curves converge to a constant value. For
higher $z$ it starts to decrease.  This means that the method is ambiguous: for
the given value of PFR one may have two redshifts - below and above $z \approx
4$.  Because for the bright GRBs the values above $z \approx 4$ are practically
excluded, for them the method is usable.  Using only the $25-55$ keV and
$55-100$ keV BATSE energy channels, this method can be used to estimate GPZ
only in the redshift range $z\; ^{<}_{\sim}\; 4$.

\vskip1mm
\begin{wrapfigure}{2}[0pt]{61mm}
\centerline{\psfig{figure=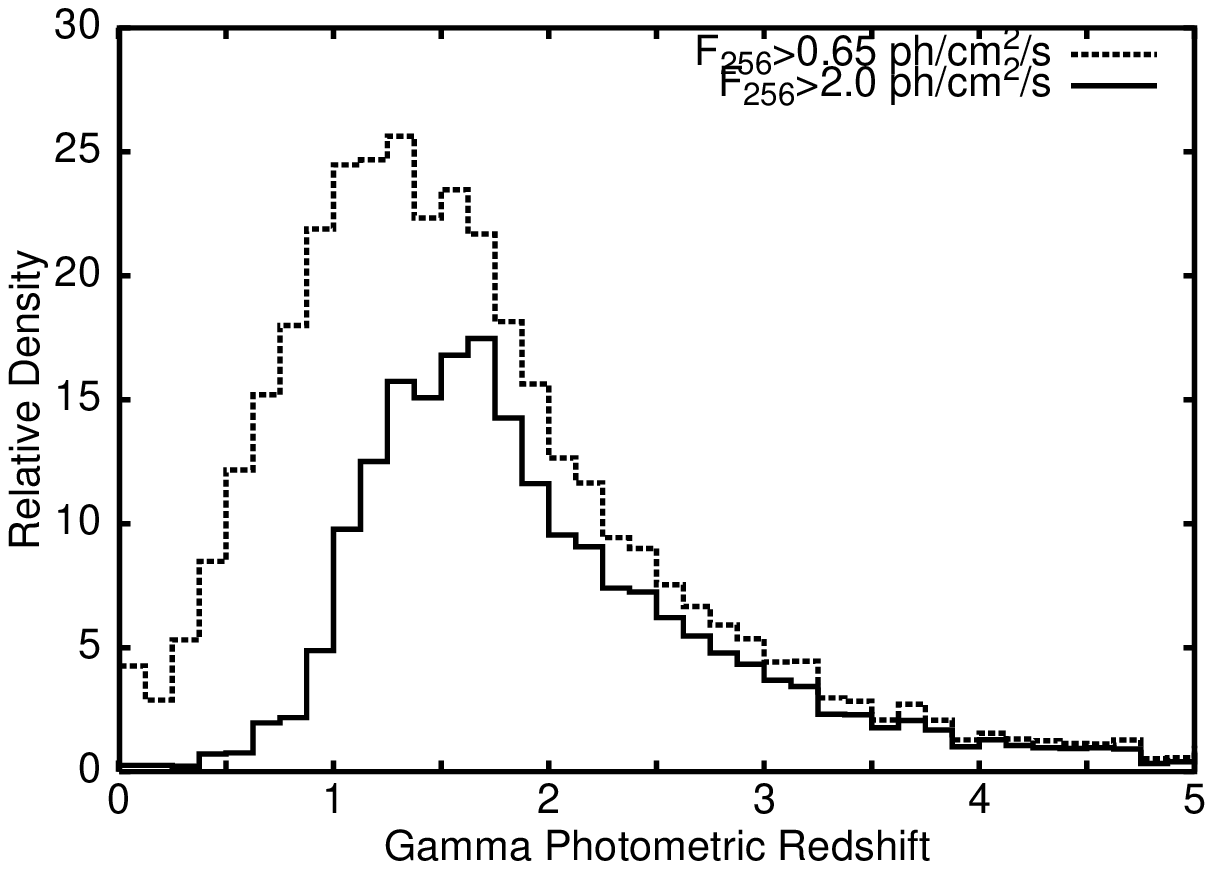,width=60truemm,angle=0,clip=}}
\captionb{1}{ The distribution of the GPR estimators
of the long GRBs having DISCSC data.  }
\end{wrapfigure}

Let us assume for a moment that all observed long bursts, we have selected
above,  have $z < 4$.  Then we can simply calculate the $z^{GPZ}$ redshift for
any GRB, which has PFR from the DISCSC data.  Fig.~2. shows the
distribution of the estimated derived redshifts {\it under the assumption that
all GRBs are below} $z \approx 4$.  The distribution has a clear peak value
around PFR $\approx 0.2$, which corresponds to $z \approx (1.5- 2.0)$.

Although there is a problem with the degeneracy (e.g. two possible redshift
values) we think that the great majority of values of $z$ obtained for the
bright half are correct.  This opinion may be supported by the following
arguments: the obtained distribution of GRBs in $z$ for the bright half 
is very similar to the obtained distribution of {Schmidt
(2001)} and Schaefer {\em et. al} (2001).  An another problem for $z$  as it
moves into $z ^{>}_{\sim} 4$ regime for the bright GRB is the extremely high
GRB luminosities,  $\simeq 10^{53} ergs/s$ (M\'esz\'aros \& M\'esz\'aros, 1996).

\vskip1mm
\begin{wrapfigure}{i}[0pt]{61mm}
\centerline{\psfig{figure=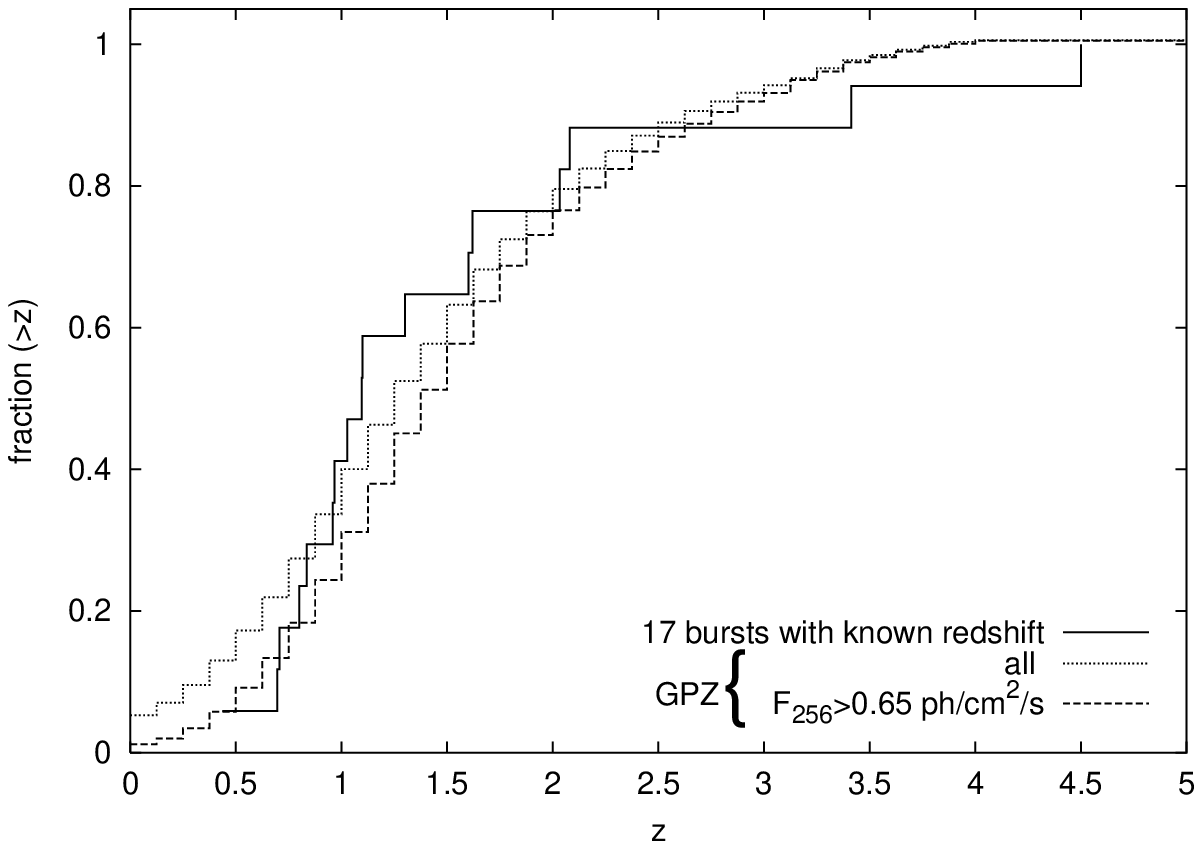,width=65truemm,angle=0,clip=}}
\captionb{3}{
The redshift distribution of the 17 GRBs' with known $z$
and the distributions from the GPZ estimators.
}
\end{wrapfigure}
As an additional statistical test we compared the redshift distribution of
the 17 GRB with observed redshift with our reconstructed GRB $z$ distributions
(limited to the $z<4$ range).  For the  $F_{256} > 0.65$ photon/(cm${}^2$s) group
the KS test suggests a $38\%$ probability, i.e. the observed
$N(<z)$ probability distribution agrees quite well with the GPZ reconstructed
function. 

\smallskip

ACKNOWLEDGMENTS\\
The useful remarks with Drs. T. Budav\'ari, S. Klose, D. Reichart, A.S.  Szalay are kindly acknowledged.
This research was supported in part through OTKA grants T024027 (L.G.B.), F029461 (I.H.) and T034549,
Czech Research Grant J13/98: 113200004 (A.M.), NASA grant NAG5-9192 (P.M.).
\goodbreak
\References

\ref Budav\'ari, T., Csabai, I., Szalay, A.S. {\em et. al},  2001,  AJ, 122, 1163

\ref Csabai, I., Connolly, A.J., Szalay, A.S. {\em et. al}, 2000, AJ, 119, 69

\ref Hakkila, J., Haglin, D. J., Pendleton, G. N. {\em et. al},  2000, ApJ, 538, 165

\ref Klose, S. 2000, Reviews in Modern Astronomy 13, Astronomische Gesellschaft, Hamburg, p.129

\ref M\'esz\'aros, A., \& M\'esz\'aros, P. 1996, ApJ, 466, 29

\ref Preece, R.D., Briggs, M.S., Pendleton, G.N., {\em et. al}  1996, ApJ, 473, 310

\ref Preece, R.D., Briggs, M.S., Mallozzi, {\em et. al},  2000, ApJS, 126, 19

\ref Ryde, F., \& Svensson, R. 1999, ApJ, 512, 693

\ref Ryde, F., \& Svensson, R. 2000, ApJ, 529, L13

\ref Schaefer, B. E., Deng, M. \& Band, D. L., 2001, ApJ, 563, L123

\ref Schmidt, M. 2001, ApJ, 552, 36

\end{document}